\documentclass[aps,preprint,showpacs,preprintnumbers]{revtex4}
\usepackage{graphicx}
\usepackage{dcolumn}
\usepackage{bm}
\def\p{"particle" $\;$}
\def\gw{gateway state $\;$}
\def\gws{gateway states $\;$}
\def\es{environmental state $\;$}
\def\ess{environmental states $\;$}
\def\rs{remote state $\;$}
\def\rss{remote states $\;$}

\def\ga{$\mid g_\alpha\rangle$}
\def\gb{$\mid g_\beta\rangle$}
\def\wa{$\mid w_\alpha\rangle$}
\def\wb{$\mid w_\beta\rangle$}
\def\ka{$\mid \kappa_\alpha\rangle$}

\def\ktm{\langle\tilde{\kappa}-\mid}

\def\ktp{\langle \tilde{\kappa}+\mid}
\def\ktpm{\langle \tilde{\kappa}\pm\mid}

\def\hktm{\ktm H\mid g_\alpha\rangle}
\def\hktp{\ktp H\mid g_\alpha\rangle}

\def\sint{{\rm sin (\Delta\varepsilon\it t/\hbar)}}
\def\sinttt{\frac{\rm sin (3\Delta\varepsilon\it t/\hbar)}{3}}
\def\sinvt{\frac{\rm sin (5\Delta\varepsilon\it t/\hbar)}{5}}

\def\s{$\;$}
\def\bs{\hspace*{-.25cm}}

\begin{document}

\title{Telegraph signals as a solution of the time dependent Schr\"odinger equation}

\author{D. Drakova}
\affiliation{University of Sofia, Faculty of Chemistry, Sofia, Bulgaria}
\email{drakova@inorg.chem.uni-sofia.bg}
\author{G. Doyen}
\affiliation{Ludwig-Maximilians University, Munich, Germany}
\homepage{http://www.cenat.de/}

\begin{abstract}
A particle switching between two sides of a symmetric system in interaction with a continuum  
exhibits a telegraph-like time development without the need of 
the Born-Bohr principle of reduction on eigenstates of the measuring 
equipment. The origin of the telegraph signal is a very weak local
coupling of the particle to the continuum which is connected with an enormous
slow down of the particle motion. The proposed mechanism
might serve as a useful simple model for studying decoherence effects due to coupling to the 
environment.
\end{abstract}

                              
\maketitle

\section{Introduction}
Telegraph signals have been observed in experiments of very different 
nature. They are displayed as sudden changes of a measured signal between two or more values
with time, appearing in a random, statistical manner (cf. for instance 
\cite{mosfetnoise}-\cite{nacci}). \\

\noindent The common theoretical approaches to random telegraph signals 
involve stochastic elements and 
range from hidden Markov models, to Marcus theory and classical or quantum 
stochastic approaches \cite{suarez} (more details are found in ref. \cite{parisi-auletta}).
The Bohr-Born concept of state reduction \cite{bohr-born} 
is implied by the state-of-the-art decoherence theory
where permanent ''measurement'' by the environment of a local system
is assumed to lead to special pointer states,
into which the system collapses. 
Within collapse theories 
the explanation of the statistical change of the state of a measured local system
relies on truncation of the coherent time development,
as it is described with the time dependent Schr\"odinger equation. 
Between two collapses the local system evolves coherently,
at random time intervals it collapses on one or another
eigenstate of the ''measuring'' environment. 
There is no approach, to our knowledge, based on a pure 
coherent time development as Schr\"odinger's equation requires,
leading to a coherent telegraph-signal-like behaviour of a quantum system.\\

\noindent The question is: can a coherent quantum mechanical description
of a local system in interaction with the environment 
lead to the telegraph signal. In the present paper
we give a positive answer to this question. We demonstrate that it is possible for 
a system, comprising a local part, entangled to the continua of the
environmental excitations, to change state suddenly in an apparently statistical fashion,
which comes out of the solution of the
time dependent Schr\"odinger equation. The conditions   
for a telegraph-signal-like change of state 
of the system are weak and local coupling to a continuum of 
environmental excitations exhibiting soft modes.
The time development is coherent and deterministic in the 
phase space of the total system including
the environmental excitations. Focusing on the time development of the 
local system alone, which is assumed to be accessible to measurement in experiment,
it appears as if it changes state in a random way.

\vspace{.5cm}

\noindent In the next section we define the model and the Hamiltonian. Next the 
method of solution is outlined, followed by presenting the results and their discussion. 

\section{Hamiltonian}
The model describes an object called \p which moves in 
a universe consisting of two sides, side $\alpha$ and side $\beta$.
We want to investigate how long does it take for the \p to get from one
side to the other, i.e. the dynamics of side change. In the present article we
specify neither the \p nor the physical nature of the environmental
excitations. ''Particles'' like an electron or an adsorbed atom and the 
environmental continua of electron-hole pairs, phonons, etc. may be envisioned.  

\vspace{.5cm}

\noindent On each side of the universe we have the same number and kind of states
in which the \p can reside:
\begin{enumerate}
\item One \rs \ga\s and \gb , respectively.
From these \rss the \p can get to the
\gws  \wa\s on the same side or \wb\s on the other side.
\item One \gw \wa\s or \wb .
From these \gws the \p can either get
to the \ess $\{\mid \kappa_{\alpha} \rangle \}$ 
or $\{\mid \kappa_{\beta} \rangle \}$ on the same side
or to the \rs \ga\s on the same side or on the opposite side \gb .
\item $N$ \ess $\{ \mid \kappa_{\alpha}\rangle \}$ or $\{ \mid \kappa_{\beta}\rangle \}$. The number $N$
is large compared to unity and limited only by the
computational power. From the \ess
the \p can only get back to the \gw on the same side.
\end{enumerate}

\begin{figure}{} \hspace{2cm}\begin{center} \begin{minipage}{11cm}
\scalebox{0.4}{\includegraphics*{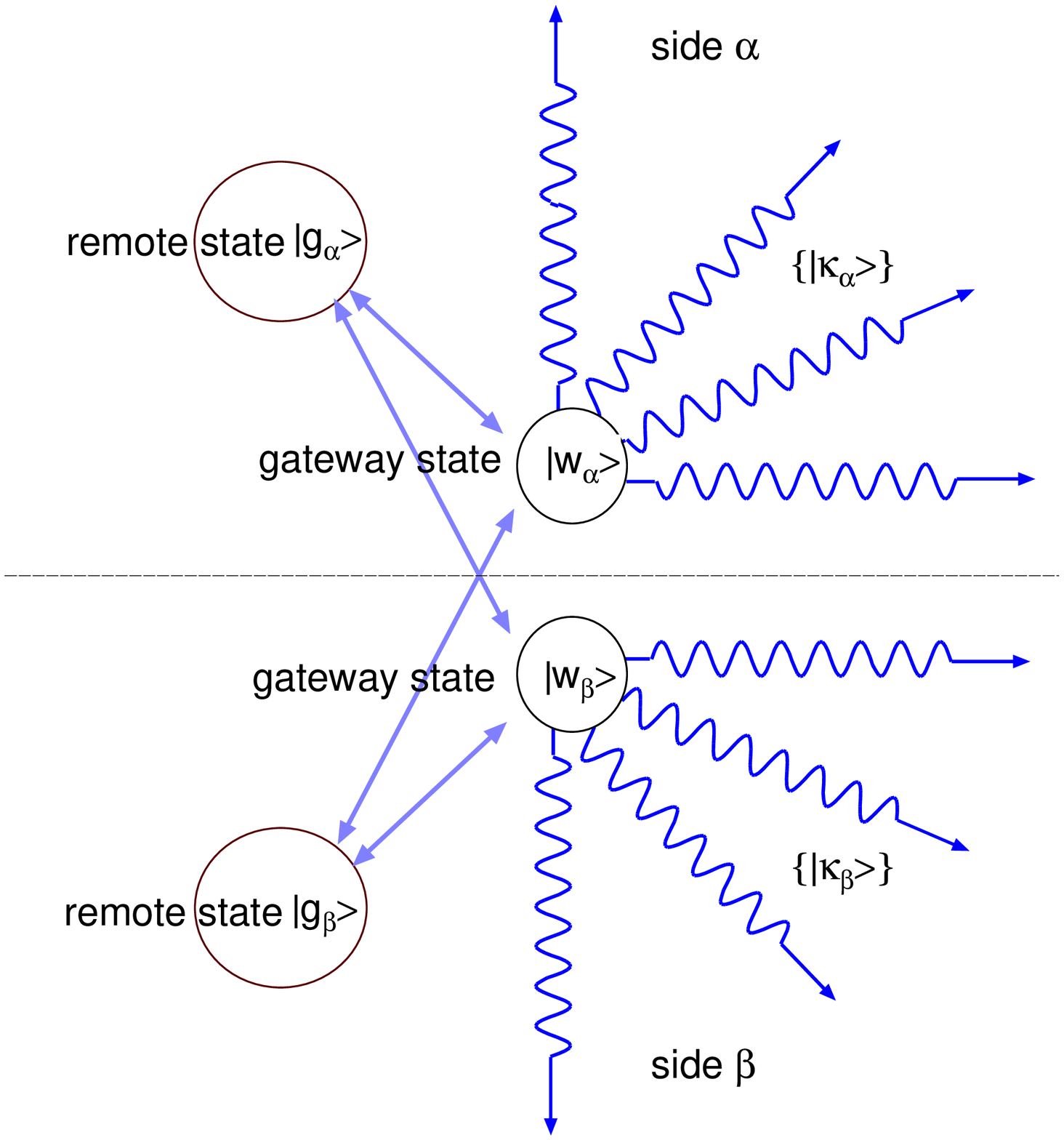}}
\end{minipage} \end{center}
\caption{Model of two \rss of a \p \ga , \gb , 
coupling within two \gws \wa , \wb\s to the \ess  $\{\mid \kappa_{\alpha} \rangle \}$, 
$\{\mid \kappa_{\beta} \rangle \}$.
\label{twosites-twowarps}} \end{figure}

\noindent The explicit form of the Hamiltonian is: 
\begin{eqnarray} \label{hamiltonian}
H&=&H_{\alpha}+H_{\beta}+H_{\alpha -\beta}\nonumber \\
H_{\alpha}&=&E_{g}n_{g\alpha} + E_{w}n_{w\alpha}
+\sum_{\kappa} \varepsilon_{\kappa} n_{\kappa \alpha}
+V(c_{g\alpha}^{+}c_{w\alpha}+h.c.) \nonumber \\
&+& W\sum_{\kappa} (c^{+}_{\kappa \alpha}c_{w\alpha}+ h.c.) \nonumber \\
H_{\beta}&=&E_{g}n_{g\beta} + E_{w}n_{w\beta}
+\sum_{\kappa} \varepsilon_{\kappa} n_{\kappa \beta}
+V(c_{g\beta}^{+}c_{w\beta}+h.c.) \nonumber \\
&+& W\sum_{\kappa} (c^{+}_{\kappa \beta}c_{w\beta}+ h.c.) \nonumber \\
H_{\alpha -\beta} &=&
(V-\Delta V)(c_{g\alpha}^{+}c_{w\beta}+c^{+}_{g\beta}c_{w\alpha}+h.c.) 
\end{eqnarray}
where $E_g$ and $E_w$ are the single-particle energies 
of the degenerate \rss \ga , \gb\s and the \gws \wa , \wb\s, respectively.
$\varepsilon_{\kappa}$ are the energies of the environmental states.
$n_{g\alpha}$ and $n_{w\alpha}$ are occupation number
operators for the \p in the states \ga , \wa ;
$c^{+}_{g\alpha}, \; c_{g\alpha}, \; c^{+}_{w\alpha}, \; c_{w\alpha}$: 
creation and destruction operators for the \p in the respective states;
$c^{+}_{\kappa \alpha}$ and $c_{\kappa \alpha}$: 
creation and destruction operators for the \p in the \ess coupling
to the gateway states; $V$ and $V-\Delta V$: interaction matrix element between
the \rss and the gateway states; $W$: interaction matrix elements between a \gw
and the \ess it couples to. 
The states $\{ \mid \kappa_{\alpha}\rangle \}$, $\{ \mid \kappa_{\beta}\rangle \}$ 
constitute a quasi-continuum which is 
modified due to entanglement to the states \wa , \wb .
The coupling of the \p to the \ess
is weak and it is mediated by the \gws\bs .\\

\subsection{\label{timedevelopment}Solution method: coherent time evolution of a \p entangled 
with a continuum of \ess }
The time evolution of the \p is according to
Schr\"odinger's time-dependent equation:
\begin{equation}
{\rm i}\hbar \frac{{\rm d} \Psi(t)}{{\rm d}t}=H\Psi(t)
\end{equation}
with $\Psi$ an eigenfunction of the total Hamiltonian eq. (\ref{hamiltonian}).
The eigenfunctions $\Psi_I$, determined by an expansion
in the basis states,
are used to derive the time dependence
according to the unitary time evolution:
\begin{equation}
\Psi_I(t) = \Psi_I(t_0)e^{-iE_{I}t/\hbar }
\end{equation}

\vspace{.5cm}
\noindent Assume that the \p starts in $\mid g_{\alpha} \rangle$.
The \p state, expanded 
in the eigenfunctions $\{ \mid I \rangle \}$ of the total Hamiltonian at time $t_0=0$ is:
\begin{equation} \label{start}
\mid g_{\alpha}(t_0)\rangle = c_{1,g_{\alpha}}\mid 1 \rangle+c_{2,g_{\alpha}}\mid 2 \rangle + ...
\end{equation}
where $c_{I,g_{\alpha}}$ is the linear coefficient of the remote state 
$\mid g_{\alpha} \rangle$ in the $I$-th eigenfunction of the Hamiltonian. 
The time development of $\mid g_{\alpha} \rangle$ is obtained as:
\begin{equation} \label{develop}
\mid g_{\alpha}(t_1) \rangle = c_{1,g_{\alpha}}e^{-iE_1t_1/\hbar}\mid 1 \rangle
+c_{2,g_{\alpha}}e^{-iE_2t_1/\hbar}\mid 2 \rangle + ...
\end{equation}
The density operator, represented in the input basis, at time $t_1$ is:
\begin{equation}
\rho(t_1) = \mid g_{\alpha}(t_1)\rangle \langle g_{\alpha}(t_1)\mid
\end{equation}
Repeating this procedure for each next time step provides 
the coherent time evolution of the density operator 
(in the input basis) which can
be used to illustrate the state changes of the \p
in interaction with the continuum of \ess\bs .

\section{Results}
\begin{figure}{} \begin{center} 
\hspace*{-5cm}
\begin{minipage}{11cm}
\scalebox{0.60}{\includegraphics*{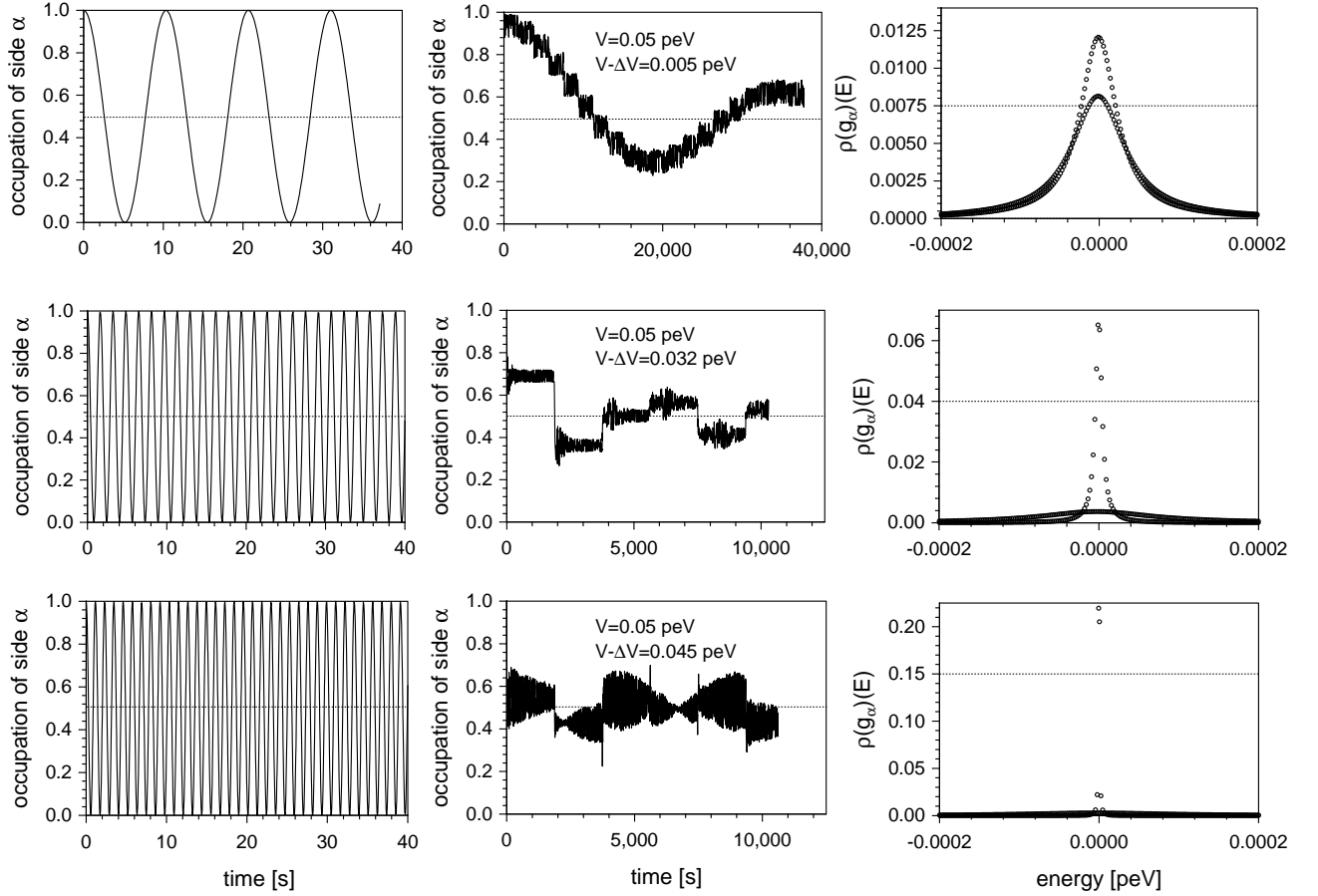}}
\end{minipage} \end{center}
\caption{Left panels: Coherent Rabi oscillations of the ''particle''  between 
the two sides $\alpha$ and $\beta$ 
in fig. \ref{twosites-twowarps} determine the side variation 
with time. No entanglement to \ess within the
\gws is included. The occupation of side $\alpha$ (eq. \ref{occga}) 
is plotted as a function of time.\\
\noindent Central panels: Switching of the \p between 
the sides $\alpha$ and $\beta$ results with
the entanglement to the \ess $\{ \mid \kappa_{\alpha} \rangle \}$ and 
$\{ \mid \kappa_{\beta} \rangle \}$ within the \gws \wa , \wb\s taken into account.
The time development of the occupation of side $\alpha$ eq. (\ref{occgalpha-envir}) is plotted.\\
\noindent Right panels: Spectral distribution of $\mid g_{\alpha}\rangle$ 
within the band of environmental states (eq. \ref{spdistr}). 
The telegraph signal in the middle central panel is 
provided by the time development of the component $\mid g-\rangle$ (fig. \ref{levelscheme}).
\label{nograv-grav-spec}} 
\end{figure}

We provide three numerical examples with 800 continuum states. 
The parameters are summarized in table \ref{examples}.
Results of these calculations are plotted in fig. \ref{nograv-grav-spec}.\\
\begin{center}
\begin{table}[h]{}  
\begin{minipage}{15cm}
\caption{\label{examples}Parameter values used for the results displayed in fig. 
\ref{nograv-grav-spec}.}
\vspace*{0.5cm} 
\begin{tabular}{|l|c|c|c|c|c|c|} 
\hline
example    & $E_g$ [peV] & $E_w$ [peV] & $\Delta \varepsilon$ [peV] & $V$ [peV] & $W$ [peV]& $\Delta V$ [peV]\\
\hline
1 example & 0.00  & 2.50 & $2.22 \times 10^{-6}$ & 0.05 & 0.00707 & 0.045 \\
\hline
2 example & 0.00  & 2.50 & $2.22 \times 10^{-6}$ & 0.05 & 0.00707 & 0.018 \\
\hline
3 example & 0.00  & 2.50 & $2.22 \times 10^{-6}$ & 0.05 & 0.00707 & 0.005  \\
\hline
\end{tabular}
\end{minipage}
\end{table}
\end{center}

\noindent The time development of the system without entanglement to
the \ess (i.e. $W=0$) is displayed in the left panels of fig. \ref{nograv-grav-spec} for
the three examples.
The quantity plotted is the occupation of side $g_{\alpha}$: 
\begin{equation} \label{occga}
\langle g_{\alpha}\mid \rho(t) \mid g_{\alpha}\rangle+
\langle w_{\alpha}\mid \rho(t) \mid w_{\alpha}\rangle 
\end{equation}
as a function of time.

\vspace{.5cm}
\noindent The system starts on the $\alpha$-side with the \p
occupying the state \ga\s.  
The time development of the system
is the result of solving Schr\"odinger's time dependent
equation as it is described in the previous section \ref{timedevelopment} and in refs. 
\cite{acsinpaper,dicepapers}. 
Within a certain time interval, which depends on the coupling strength
between the states $\mid g_{\alpha} \rangle$, $\mid g_{\beta} \rangle$ and
$\mid w_{\alpha} \rangle$, $\mid w_{\beta} \rangle$,
the occupation of side $\alpha$ reduces and the 
particle changes side to $\beta$. The stronger the coupling between the remote states
and the gateway states is, the faster the particle changes from one side to the other. 
It is obvious though, that the changes of the side occupation with time are sine-like, rather 
than telegraph-signal-like. These are just coherent    
Rabi oscillations between the two sides \ga\s 
and \gb, mediated by coupling to the
\gws $\mid w_{\alpha} \rangle$ and \wb . The frequency of side transition is determined by the energy
difference $\Delta \varepsilon$ between the eigenstates of the system involving 
the \rss and the \gws only: \ga, \gb, \wa, \wb .

\vspace{.5cm}
\noindent Only with entanglement to the \ess included, does a telegraph-signal-like
time developement of the side switching arise as it is seen 
on the central panel of fig. \ref{nograv-grav-spec}, 
where the changes of the occupation of side $g_{\alpha}$ are
plotted as a function of time:
\begin{equation} \label{occgalpha-envir}
\langle g_{\alpha}\mid \rho(t) \mid g_{\alpha}\rangle
+\langle w_{\alpha}\mid \rho(t) \mid w_{\alpha}\rangle+
\sum_{\kappa \alpha}\langle\kappa_{\alpha}\mid \rho(t)\mid\kappa_{\alpha} \rangle
\end{equation}
The telegraph-like change of side is observed only in the 
second row, central panel with the choice of the coupling
parameter between the remote states and the gateway states $V-\Delta V=0.032$ peV.

\vspace{.5cm}
\noindent The eigenstates $\{\mid I \rangle \}$ are obtained by diagonalizing
numerically a $800 \times 800$-matrix representing the Hamiltonian.
An apparently statistical telegraph-like
change of side of the \p is observed in the second example, displayed in the middle central panel
in fig. \ref{nograv-grav-spec}. 
Furthermore, a significant slow down of the switching rate
between the sides results,
compared to the period of the Rabi oscillation, neglecting the entanglement to \ess
(cf. fig. \ref{nograv-grav-spec}, left and central panels). In the telegraph regime  
the average frequency of the telegraph signal is determined by a much smaller energy 
difference $\Delta \varepsilon = 2.2\times 10^{-6}$ peV equal to the energy separation of the \ess
which entangle to the \p movement.

\section{Discussion \label{discussion}}

\subsection{Symmetry adapted basis states}
\begin{figure}{}\begin{center} \begin{minipage}{11cm}
\scalebox{0.5}{\includegraphics*{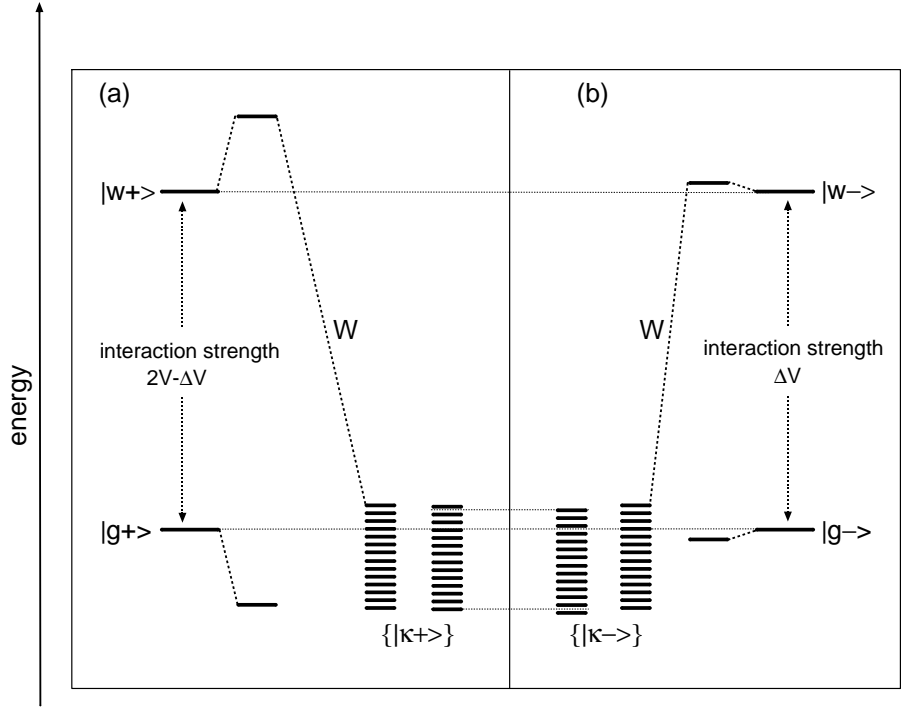}}
\end{minipage} \end{center}
\caption{Schematic energy levels
resulting from coupling between the transformed
basis states, belonging to the $\{ + \}$ (a) and $\{ - \}$ (b) subsystems 
(eqs. \ref{gw-transf} and \ref{kappa-transf}).
The degeneracy between the transformed continua 
$\{\mid \kappa+\rangle \}$ and $\{\mid \kappa-\rangle \}$
is lifted due to coupling to $\mid w+\rangle$ (a) and to $\mid w-\rangle$ (b), respectively.
The \p in the $\mid g- \rangle$-component alone stays approximately on-shell with the  
environmental states because of the weaker interaction of the order of $\Delta V$ with $\mid w- \rangle$.
\label{levelscheme}} \end{figure}

The model system is obviously symmetric with respect to interchanging the
indices $\alpha$ and $\beta$, i.e., interchanging the two sides.
The eigenstates are therefore either symmetric or anti-symmetric with respect
to such an interchange.
We therefore make the following transformation:
\begin{eqnarray} \label{gw-transf}
c_{g+}&=&\frac{1}{\sqrt{2}}(c_{g \alpha}+c_{g \beta}) \nonumber \\
c_{g-}&=&\frac{1}{\sqrt{2}}(c_{g \alpha}-c_{g \beta}) \nonumber \\
c_{w+}&=&\frac{1}{\sqrt{2}}(c_{w \alpha}+c_{w \beta}) \nonumber \\
c_{w-}&=&\frac{1}{\sqrt{2}}(c_{w \alpha}-c_{w \beta})
\end{eqnarray}
The coupling matrix elements are then:
\begin{eqnarray}
\langle g_+ \mid H \mid w_+ \rangle&=&\frac{1}{2}(4V-2\Delta V)= 2V-\Delta V \nonumber \\
\langle g_+ \mid H \mid w_- \rangle&=&\frac{1}{2}(V-V-V+\Delta V+V-\Delta V)=0 \nonumber \\
\langle g_- \mid H \mid w_+ \rangle&=&0 \nonumber \\
\langle g_- \mid H \mid w_- \rangle&=&\frac{1}{2}(V+V-V+\Delta V -V +\Delta V)=\Delta V
\end{eqnarray}
Transformed operators for \ess are defined as:
\begin{eqnarray} \label{kappa-transf}
c_{\kappa+}&=&\frac{1}{\sqrt{2}}(c_{\kappa \alpha}+c_{\kappa \beta}) \nonumber \\
c_{\kappa-}&=&\frac{1}{\sqrt{2}}(c_{\kappa \alpha}-c_{\kappa \beta}) 
\end{eqnarray}
The coupling matrix elements between the \gws and the \ess  are then:
\begin{eqnarray}
\langle w_+ \mid H \mid \kappa_+ \rangle&=&\frac{1}{2}(W+W)= W \nonumber \\
\langle w_- \mid H \mid \kappa_+ \rangle&=&0 \nonumber \\
\langle w_+ \mid H \mid \kappa_- \rangle&=&0 \nonumber \\
\langle w_- \mid H \mid \kappa_- \rangle&=&\frac{1}{2}(W+W)=W
\end{eqnarray}
With these transformed operators the Hamiltonian reads:
\begin{eqnarray} \label{transfhamiltonian}
H&=&H_++H_-\nonumber \\
H_+ &=&E_{g}n_{g+} + E_{w}n_{w+}+\sum_{\kappa} \varepsilon_{\kappa}n_{\kappa+} \nonumber \\
&+&(2V-\Delta V)(c_{g+}^{+}c_{w+}+c^{+}_{w+}c_{g+})+ W\sum_{\kappa} (c^{+}_{\kappa+}c_{w+}+c^{+}_{w+}c_{\kappa+}) \nonumber \\
H_- &=&E_{g}n_{g-}+E_{w}n_{w-}+\sum_{\kappa} \varepsilon_{\kappa}n_{\kappa-} \nonumber \\
&+&\Delta V(c_{g-}^{+}c_{w-}+c^{+}_{w-}c_{g-})+ W\sum_{\kappa} c^{+}_{\kappa-}c_{w-} + c^{+}_{w-}c_{\kappa-}) 
\end{eqnarray}
The consequence is that the $\{ \mid \kappa+\rangle \}$ and $\{ \mid \kappa-\rangle \}$ systems are decoupled. The mixing
between the \rs\s and the on-shell environmental states is weaker in the antisymmetric system, if $\Delta V$
is significantly smaller than $(2V-\Delta V)$, i.e. in the telegraph and bonding regimes
(see below).
The interactions in each system are illustrated in fig. \ref{levelscheme} for this situation.
The degeneracy between the continua 
$\{\mid \kappa+ \rangle \}$ and $\{\mid \kappa-\rangle \}$
is lifted because of the different \rs - \gw coupling in the $\{ + \}$ and $\{ - \}$ subsystems. 
If $\Delta V$ is small and the time development of the \p starts 
from \ga\s then the \p in the
$\mid g- \rangle$-component alone stays approximately on-shell and, due to the weakness
of the interaction $W$, mixes significantly with the on-shell environmental states.
The energy shell is defined by the energy of the \p in its initial state $\mid g_{\alpha} \rangle$.
This energy is set equal to zero.
This means that \ga\s mixes in only with the $\{ \mid \kappa -\rangle \}$ continuum 
because $\{ \mid \kappa+\rangle \}$ and $\{ \mid \kappa-\rangle \}$ states are not mixed and only states belonging to $\{ \mid \kappa -\rangle \}$
are nearly on-shell with $\mid g_{\alpha} \rangle$.

\subsection{Coupling regimes}
The characteristics of the time development of a wave packet starting in the \p state
\ga\s are determined by
the Hamiltonian matrix element between the initial state $\mid g_\alpha \rangle$
and the eigenstates $\{\mid \tilde{\kappa}\pm\rangle\}$ derived from the \ess\bs .  
This matrix element might be
estimated as follows, if the coupling between the gateway state \wa\s 
and the \es\s \ka\s is weak enough so that
perturbation theory can be used to calculate the coefficients $\ktpm w \pm\rangle$:
\begin{eqnarray} \label{contcoupling1}
\hktm &\approx&\ktm H\mid w-\rangle\langle w-\mid g_\alpha\rangle 
+\ktm w-\rangle\langle w-\mid H\mid g_\alpha\rangle
\nonumber \\
&\approx &W\times 0 +\frac{W}{E_w}\frac{1}{\sqrt{2}}\Delta V 
\end{eqnarray}
\begin{eqnarray} \label{contcoupling2}
\hktp &\approx&\ktp H\mid w+\rangle\langle w+\mid g_\alpha\rangle 
+\ktp w+\rangle\langle w+\mid H\mid g_\alpha\rangle
\nonumber \\
&\approx &W\times 0 +\frac{W}{E_w}\frac{1}{\sqrt{2}}(2V-\Delta V)
\end{eqnarray}

\vspace{.5cm}

\noindent 
These estimates serve only to define
regimes of different coupling strength of \ga\s to the continuum.
In table \ref{regimes} ranges of values of the coupling parameters
between the initial particle state $\mid g_{\alpha} \rangle$ and
the environmental states belonging to the two continua 
$\{ +  \}$ and $\{ -  \}$ are indicated. The second column
includes intervals for the values of $\Delta V$, proportional to the coupling strength to 
the $\{ \mid {\tilde \kappa}- \rangle \}$ states. In the third column 
intervals for $2V-\Delta V$, proportional to the coupling 
strength to $\{\mid  {\tilde \kappa}+ \rangle \}$, are summarized.  

\begin{center}
\begin{table}[h]{}  
\begin{minipage}{15cm}
\caption{\label{regimes}Regimes for different kinds of dynamics of a particle starting in 
state $\mid g_{\alpha} \rangle$, in interaction with the
continua of environmental states $\{ \mid {\tilde \kappa}- \rangle \}$ 
and $\{ \mid {\tilde \kappa}+ \rangle \}$.}
\vspace*{0.5cm} 
\begin{tabular}{|l|c|c|} 
\hline
regime & $\hktm\frac{\sqrt{2} E_w}{W}=\Delta V$ & $\hktp\frac{\sqrt{2}E_w}{W}=2V-\Delta V$\\
\hline
slow Rabi oscillations & $V \; ... \;  \frac{5}{6}V$ & $V  \; ... \;  \frac{7}{6}V$\\
\hline
faster  Rabi oscillations & $\frac{5}{6}V  \; ... \;  \frac{1}{2}V$ & $\frac{7}{6}V \;  ... \;  \frac{3}{2}V$\\
\hline
telegraph & $\frac{1}{2}V  \; ...  \; \frac{1}{6}V$ & $\frac{3}{2}V \;  ...  \; \frac{11}{6}V$ \\
\hline
bonding & $\frac{1}{6}V  \; ...  \; 0$ & $\frac{11}{6}V \;  ...  \; 2V$    \\
\hline
\end{tabular} 
\end{minipage}
\end{table}
\end{center}

\vspace{.5cm}

\noindent 
Illustrations of these regimes are displayed in the three horizontal panels
in fig. \ref{nograv-grav-spec} for three different choices of the parameter $V-\Delta V$.
In the regime of slow Rabi oscillations the hopping matrix element in the Hamiltonian
$V-\Delta V$ for switching between the $\alpha$- and the $\beta$-side
is near zero. \ga\s couples with approximately equal strength $V$, which
is rather large, 
to the $\{ \mid {\tilde \kappa}+\rangle \}$  and 
$\{ \mid {\tilde \kappa}-\rangle \}$ continua. The spectral distribution of \ga\s
in the two continua is defined as:
\begin{equation} \label{spdistr}
\rho_{g\alpha}(E) = \sum_{\tilde{\kappa}\pm}\mid\ktpm g_\alpha\rangle \mid^2\delta (E-E_{\kappa\pm}) 
\end{equation}
It is rather broad near the energy shell
and exhibits approximately equal weights in the two continua
(cf. fig. \ref{nograv-grav-spec}, upper right panel). The \p
oscillates slowly between the two sides (fig. \ref{nograv-grav-spec}, upper central panel).
The long oscillation time is determined by the off-shell environmental states, 
whose degeneracy is only very slightly lifted because the coupling
to both environmental continua $\{ \mid {\tilde \kappa} -\rangle \}$
and $\{ \mid {\tilde \kappa} +\rangle \}$ is nearly equally strong. The slow Rabi oscillation
is superposed by telegraph-like jumps, the jump time being determined by
the the value of $\Delta \varepsilon$ near the energy shell.

\vspace{.5cm}

\noindent 
In the regime of faster Rabi oscillations, $V-\Delta V$ deviates significantly from zero.
The coupling of \ga\s to the $\{ \mid {\tilde \kappa}+\rangle \}$ continuum
is already significantly stronger than to the $\{ \mid {\tilde \kappa}-\rangle \}$ continuum. The spectral weight
of \ga\s in the $\{ \mid {\tilde \kappa}+\rangle \}$ continuum is therefore shifted from the on-shell region
to lower energies and the spectral weight near the energy shell is significantly smaller
than in the $\{ \mid {\tilde \kappa}-\rangle \}$ continuum. No illustration of this regime is given in this paper.

\vspace{.5cm}

\noindent 
In the region where the telegraph-like switching between sides occurs, 
the difference between the coupling to the $\{ \mid {\tilde \kappa}+\rangle \}$ 
and $\{ \mid {\tilde \kappa}-\rangle \}$
continua increases. The coupling to the $\{ \mid {\tilde \kappa}-\rangle \}$ continuum becomes
now so small that the spectral distribution of \ga\s in the $\{ \mid {\tilde \kappa}-\rangle \}$ continuum
is represented by a narrow Lorentzian (fig. \ref{nograv-grav-spec}, middle right panel)
responsible for the telegraph signal in the middle central panel, as explained below.
The long time transitions are no more discernible because the time development
is dominated by the on-shell eigenstates derived from environmental states
in the $\{ \mid {\tilde \kappa}-\rangle \}$ continuum. This yields the telegraph-like behaviour.

\vspace{.5cm}

\noindent 
In the bonding regime $V-\Delta V$ equals aproximately $V$
and the coupling between the $\alpha$ and $\beta$ sides
is so strong that \ga\s and \gb\s form a linear combination similar to
a chemical bond (fig. \ref{nograv-grav-spec}, low right panel). 
The \p is now on the two sides at the same time.
There are then no long-time transitions between
the $\alpha$ and the $\beta$ side, though 
telegraph-like variations of the occupation of side $\alpha$ are still discernible. 
(fig. \ref{nograv-grav-spec}, low central panel).

\subsection{Diagonalization in the case of a degenerate continuum}

We first demonstrate that the telegraph signals are a consequence of the
environmental states giving rise to energetically close spaced quasi-continuum states.
In the case that all \ess are degenerate only energetically well separated
off-shell eigenstates are coupled to the \gws giving rise to high frequency
sine-like Rabi-oscillations. 

\def\wktm{\mid w_{\kappa} -\rangle}
\def\wkgm{\mid w_{\kappa g}-\rangle}
\def\DV{$\Delta V$}

\vspace{.5cm}

\noindent 
We calculate the antisymmetric eigenstates for the degenerate \ess\bs .
Defining the projected state $\mid w_{\kappa} -\rangle$ 
\begin{equation}    \label{wkappa}
\wktm = \frac{1}{\sqrt{N}}\sum_{\kappa -}\mid \kappa -\rangle
\end{equation}
the Hamiltonian in the anti-symmetric subspace takes the form
\begin{center}
\begin{tabular}{c|ccc}
  & $\mid w-\rangle$     & $\mid w_{\kappa} -\rangle$ & $\mid g-\rangle$ \\
\hline  
$\langle w-\mid$ & $E_w$ & $\sqrt{N}W$ & \DV  \\
$\langle w_{\kappa} -\mid$ & $\sqrt{N}W$ & 0 & 0  \\
$\langle g-\mid$          & \DV & 0 & 0
\end{tabular}
\end{center}
$N$ is the number of degenerate environmental states.
This reduces to a $2\times 2$-system by defining
\begin{equation} \label{wkgm}
\wkgm = \frac{\sqrt{N}W\wktm + \Delta V\mid g-\rangle }{\sqrt{NW^2+(\Delta V)^2}}.
\end{equation}
Orthogonalizing $\mid g-\rangle$ to $\wkgm$
\begin{equation} \label{gmortho}
\mid g_{\perp}-\rangle = \sqrt{\frac{NW^2+(\Delta V)^2}{NW^2}}\left(\mid g-\rangle
-\langle w_{\kappa g}-\mid g-\rangle \wkgm\right),
\end{equation}
the Hamiltonian matrix transforms into
\def\sqwv{\sqrt{NW^2+(\Delta V)^2}}
\begin{center}
\begin{tabular}{c|ccc}
  & $\mid w-\rangle$     & $\wkgm$ & $\mid g_{\perp}-\rangle$ \\
\hline  
$\langle w-\mid$ & $E_w$ & $\sqwv$ & 0  \\
$\langle w_{\kappa g}-\mid$ & $\sqwv$ & 0 & 0  \\
$\langle g_{\perp}-\mid$         & 0 & 0 & 0
\end{tabular}
\end{center}
The eigenvalues are:
\begin{eqnarray}
E_1&=&0\nonumber \\
E_{2/3}&=&\frac{E_w}{2}\pm\frac{E_w}{2}\sqrt{1+\frac{4(NW^2+(\Delta V)^2)}{E_w^2}}.
\end{eqnarray}
$E_3$ is the energy of the $\wkgm$ derived eigenstate and is roughly equal to
\begin{equation}
E_3\approx -\frac{NW^2+(\Delta V)^2}{E_w}
\end{equation}
which, though smaller in magnitude than $E_w$, is energetically well separated
from the on-shell energy zero.

\vspace{.5cm}

\noindent 
Orthogonalizing the \ess to $\wkgm$
\begin{equation}      \label{kappaortho}
\mid\kappa_{\perp}-\rangle  =\frac{1}{\sqrt{1-\mid\langle\kappa -\wkgm\mid^2}}\left(
\mid \kappa -\rangle - \frac{W}{\sqwv}\wkgm\right),
\end{equation}
the antisymmetric states are split in two separated and orthogonal subspaces $\{\mid w-\rangle ,\;\wkgm \}$
and $\{\mid g_{\perp}-\rangle,\;\mid \kappa_{\perp}-\rangle\}$, where the latter is energetically
on-shell at energy zero and the former is energetically off-shell.
$\langle\kappa -\wkgm$ is obtained from eq. (\ref{wkgm}) as
\begin{eqnarray} \label{eq23}
\langle\kappa -\wkgm &=& \frac{\sqrt{N}W\frac{1}{\sqrt{N}}}{\sqrt{NW^2+(\Delta V)^2}} \nonumber \\
&=&\frac{W}{\sqrt{NW^2+(\Delta V)^2}}
\end{eqnarray}
which for the results of the center panel in the second row of fig. \ref{nograv-grav-spec}
is roughly $5\times 10^{-2}$.

\subsection{Interaction in the on-shell subspace}
\subsubsection{Hamiltonian matrix elements in the on-shell subspace}
We now investigate the interaction of the $\mid g_{\perp}-\rangle$-state with the
$\{ \mid\kappa_{\perp}-\rangle \}$-states for the {\em non-}degenerate environmental states continuum
using two approximations:
\begin{enumerate}
\item The $\{\mid g_{\perp}-\rangle ,\; \mid \kappa_{\perp}-\rangle\}$-states are mutually orthogonal.
\item The Hamiltonian matrix elements between the  $\{\mid w-\rangle , \;\wkgm \}$ subspace
and the  $\{\mid g_{\perp}-\rangle ,\;\mid \kappa_{\perp}-\rangle\}$ subspace vanish.
\end{enumerate}
The first assumption is violated to order $N^{-1}$, which becomes obvious by calculating
the overlap from eqs. (\ref{gmortho}) and (\ref{kappaortho}):
{\normalsize
\begin{eqnarray}
\langle\kappa_{\perp}-\mid g_{\perp}-\rangle &=&\sqrt{ \frac{NW^2+(\Delta V)^2}{NW^2} }\sqrt{ \frac{NW^2+(\Delta V)^2}{(N-1)W^2+(\Delta V)^2 } }\nonumber \\
&\times &\left( \langle\kappa -\mid g-\rangle - \frac{2W\langle w_{\kappa g}-\mid g-\rangle}{\sqrt{NW^2+(\Delta V)^2}} 
-\langle w_{\kappa g}-\mid g-\rangle\langle\kappa -\mid w_{\kappa g}-\rangle \right)
\end{eqnarray} }
where
\begin{eqnarray}
\langle\kappa -\mid g-\rangle &=& 0 \\
\langle g-\mid w_{\kappa g}-\rangle &=& \frac{\Delta V}{\sqrt{NW^2+(\Delta V)^2}}
\end{eqnarray}
and $\langle\kappa -\mid w_{\kappa g}-\rangle =\frac{W}{\sqrt{NW^2+(\Delta V)^2}}$ 
according to eq. (\ref{eq23}). \\

\vspace{.2cm}
\noindent The second assumption is violated
to order $N^{-\frac{1}{2}}\Delta\varepsilon$. So for increasing number of \ess
and decreasing energetic separation of the \ess\bs , the approximations become
arbitrarily well fulfilled.

\vspace{.5cm}

\noindent 
We need the Hamiltonian matrix element between 
$\mid g_{\perp}-\rangle$ and $\mid \kappa_{\perp}-\rangle$:
{\normalsize
\begin{eqnarray} \label{k-Hg-}
\langle\kappa_{\perp}-\mid H\mid g_{\perp}-\rangle &=& \sqrt{\frac{NW^2+(\Delta V)^2}{NW^2}}\Bigl(
\langle\kappa_{\perp}-\mid H\mid g-\rangle - \langle w_{\kappa g}-\mid g-\rangle\langle\kappa_{\perp}-\mid H\wkgm \Bigr)  \nonumber\\
&=&  \sqrt{\frac{NW^2+(\Delta V)^2}{NW^2}} \Biggl( \frac{1}{\sqrt{1-\mid\langle \kappa -\mid w_{\kappa g}-\rangle\mid^2}}
\langle w_{\kappa g}-\mid \kappa -\rangle\langle w_{\kappa g}-\mid H\mid g-\rangle\nonumber\\
&-&\langle w_{\kappa g}-\mid g-\rangle\langle\kappa_{\perp}-\mid H\wkgm \Biggr)
\end{eqnarray}
}
The Hamiltonian matrix elements appearing in this expression are evaluated as follows:
\begin{equation}
\langle w_{\kappa g}-\mid H\mid g-\rangle =\sqrt{\frac{NW^2}{NW^2+(\Delta V)^2}}\langle w_{\kappa}-\mid H\mid w_{\kappa g}-\rangle = 0
\end{equation}
{\normalsize
\begin{eqnarray} \label{k-Hwkg}
\langle\kappa_{\perp}-\mid H\mid w_{\kappa g}-\rangle &=& \frac{\sqrt{N}W}{\sqrt{(N-1)W^2+(\Delta V)^2}}
\Bigl( \langle\kappa -\mid H\mid w_{\kappa} -\rangle
-\langle w_{\kappa g}-\mid \kappa -\rangle\langle w_{\kappa} -\mid H\mid w_{\kappa g}-\rangle \Bigr)   \nonumber \\
&=&  \frac{WE_{\kappa -}}{\sqrt{(N-1)W^2+(\Delta V)^2}}
\end{eqnarray}
}
where in the last line we used eq. (\ref{wkappa}).
Inserting in eq. (\ref{k-Hg-}) yields:
\begin{equation}\label{k-]Hg-]}
\langle\kappa_{\perp}-\mid H\mid g_{\perp}-\rangle   =  -\frac{\Delta VE_{\kappa -}}{\sqrt{N(N-1)W^2+N(\Delta V)^2}}
\end{equation}
which for the results of the center panel in the second row of fig. \ref{nograv-grav-spec}
is roughly $10^{-2}\times E_{\kappa -} $.
{\normalsize
\begin{eqnarray}  \label{l-]Hk-]}
\langle\lambda_{\perp}-\mid H\mid \kappa_{\perp}-\rangle &=&\frac{NW^2+(\Delta V)^2}{(N-1)W^2+(\Delta V)^2} \nonumber \\
&\times &\Biggl( \langle\lambda -\mid H\mid \kappa -\rangle - \frac{W}{\sqrt{NW^2+(\Delta V)^2}}
\Bigl( \langle\lambda -\mid H\mid w_{\kappa g}-\rangle + \langle\kappa -\mid H\mid w_{\kappa g}-\rangle \Bigr)\Biggr)\nonumber \\
&=&\frac{NW^2+(\Delta V)^2}{(N-1)W^2+(\Delta V)^2}\left( E_{\kappa -}\delta_{\kappa\lambda} 
-\frac{W^2}{NW^2+(\Delta V)^2}\Bigl(E_{\kappa -}+E_{\lambda -}\Bigr)\right)
\end{eqnarray}
}
which for the results of the center panel in the second row of fig. \ref{nograv-grav-spec}
is roughly $E_{\kappa -}\delta_{\kappa\lambda}-\frac{1}{N}(E_{\kappa -}+E_{\lambda -)}$.
Here we have used:
\begin{eqnarray}
\langle\lambda -\mid H\mid w_{\kappa g}-\rangle &=&\frac{1}{\sqrt{NW^2+(\Delta V)^2}}
\left(\sqrt{N}W\langle\lambda -\mid H\mid w_{\kappa} -\rangle
+\Delta V\langle\lambda -\mid H\mid g-\rangle\right)\nonumber \\
&=&\frac{1}{\sqrt{NW^2+(\Delta V)^2}}
\left(\frac{\sqrt{N}W}{\sqrt{N}}\langle\lambda -\mid H\mid \lambda -\rangle
\right)\nonumber \\
&=&\frac{WE_{\lambda -}}{\sqrt{NW^2+(\Delta V)^2}}
\end{eqnarray}

\subsubsection{Projection of remote state on environmental eigenstates}
Environmental eigenstates (exact or approximate) are denoted as $\mid\tilde{\kappa}_{\perp}-\rangle$.
We evaluate
the coefficient $\langle g_{\perp}-\mid\tilde{\kappa}_{\perp}-\rangle$ using the Lippmann-Schwinger equation \cite{lippmann}:
\begin{eqnarray} \label{lippmannschwinger}
\langle g_{\perp}-\mid\tilde{\kappa}_{\perp}-\rangle&=&\langle g_{\perp}-\mid\kappa_{\perp}-\rangle \nonumber \\
&+& \sum_\lambda \langle g_{\perp}-\mid G(E=E_{\tilde{\kappa}-})\mid \lambda_{\perp}-\rangle
\langle \lambda_{\perp}-\mid H_{-}-H_o\mid \kappa_{\perp}-\rangle 
\end{eqnarray}
where $H_o$ is the last but one line of eq. (\ref{transfhamiltonian}) and the summation over $\lambda$
is over all states in the anti-symmetric on-shell continuum. $G(E)=(E-H_-)^{-1}$ is the Green operator.
Using Dyson's equation one obtains
\begin{eqnarray} \label{lippmannschwinger2}
\langle g_{\perp}-\mid\tilde{\kappa}_{\perp}-\rangle&=&\langle g_{\perp}-\mid\kappa_{\perp}-\rangle
+ \langle g_{\perp}-\mid G_o(E=E_{\tilde{\kappa}-})\mid g_{\perp}-\rangle
\langle g_{\perp}-\mid H_{-}-H_o\mid \kappa_{\perp}-\rangle \nonumber \\
&+& \sum_{\lambda ,\mu} \langle g_{\perp}-\mid G_o(E=E_{\tilde{\kappa}-})\mid g_{\perp}-\rangle
\langle g_{\perp}-\mid H_{-}-H_o\mid \lambda_{\perp}-\rangle \nonumber \\
&\times &\langle \lambda_{\perp}-\mid G(E=E_{\tilde{\kappa}-})\mid \mu_{\perp}-\rangle
\langle \mu_{\perp}-\mid H_{-}-H_o\mid \kappa_{\perp}-\rangle 
\end{eqnarray}
where $G_o(E)=(E-H_o)^{-1}$.
Eq. (\ref{lippmannschwinger}) describes the scattering of a \p incoming in $\mid\kappa_{\perp}-\rangle$
from the perturbation provided by the \rs $\mid g_{\perp}-\rangle$ and the \ess $\mid \lambda_{\perp}-\rangle$.
Obviously outgoing boundary conditions are applied, although this
is not indicated explicitly on the Green function. 
The Green function has to be evaluated at the energy of the \es
under consideration.

\vspace{.5cm}

\noindent 
The first term on the r.h.s. of eq. (\ref{lippmannschwinger2}) vanishes, 
because the $\perp$-states are assumed to be orthogonal.
In the second term on the r.h.s. of eq. (\ref{lippmannschwinger2})
the $E_ {\kappa -}$-dependence is cancelled and 
for the results of the center panel in the second row of fig. \ref{nograv-grav-spec}
the line width resulting from this term would be of order $10^{-4}\times \Delta\varepsilon\approx 10^{-10}$
which is at least four orders of magnitude too small compared to the values
obtained numerically.

\vspace{.5cm}

\noindent 
In order to estimate the third term on the r.h.s. of eq. (\ref{lippmannschwinger2})
we could replace the exact Green operator $G$ by $G_o$ so that the third term becomes:
\begin{eqnarray}
\langle g_{\perp}-\mid\tilde{\kappa}_{\perp}-\rangle&\approx &\frac{1}{E_{\kappa -}}\frac{\Delta V}{\sqrt{(N-1)W^2+(\Delta V)^2}}
\sum_\lambda E_{\lambda -}\frac{1}{NE_{\lambda -}}
(E_{\kappa -}+E_{\lambda -})    \nonumber \\
&\approx &\frac{1}{NE_{\kappa -}}\frac{\Delta V}{\sqrt{(N-1)W^2+(\Delta V)^2}}
\left(\sum_\lambda E_{\lambda -}+NE_{\kappa -}\right) \nonumber \\
&\approx &\frac{\Delta V}{\sqrt{(N-1)W^2+(\Delta V)^2}} \nonumber
\end{eqnarray}
which is independent of $E_{\kappa -}$ and for the results of the center panel in the second row of fig. \ref{nograv-grav-spec}
roughly of order $5\times 10^{-2}$.
The line width resulting from this term would be of order $10^{-2}\times \Delta\varepsilon\approx 10^{-8}$
and too small.

\vspace{.5cm}

\noindent 
Hence replacing $G$ by $G_o$ in eq. (\ref{lippmannschwinger2}) does not yield any improvement.
This estimate shows that the telegraph signals can only appear in higher order and
therefore we substitute Dyson's equation for the exact Green operator $G$ 
in the third term on the r.h.s. of eq. (\ref{lippmannschwinger2}):
{\normalsize
\begin{eqnarray}\label{dyson}
\langle \lambda_{\perp}-\mid G\mid \mu_{\perp}-\rangle &=& \langle \lambda_{\perp}-\mid G_o\mid \mu_{\perp}-\rangle\nonumber \\
&+&\langle \lambda_{\perp}-\mid G_o\mid \lambda_{\perp}-\rangle\sum_\nu\langle \lambda_{\perp}-\mid H_{-}-H_o\mid \nu_{\perp}-\rangle 
\langle \nu_{\perp}-\mid G\mid \mu_{\perp}-\rangle
\end{eqnarray}
}
The first term reproduces the previous estimate. Re-inserting repeatedly this expression on the
r.h.s. of eq. (\ref{dyson}) one obtains the infinite series
{\normalsize
\begin{eqnarray}
\langle \lambda_{\perp}-\mid G\mid \mu_{\perp}-\rangle &=& 
\langle \lambda_{\perp}-\mid G_o\mid \lambda_{\perp}-\rangle\langle \lambda_{\perp}-\mid H_{-}-H_o\mid \mu_{\perp}-\rangle 
\langle \mu_{\perp}-\mid G_o\mid \mu_{\perp}-\rangle\nonumber\\
&+&\langle \lambda_{\perp}-\mid G_o\mid \lambda_{\perp}-\rangle\sum_\nu
\langle \lambda_{\perp}-\mid H_{-}-H_o\mid \nu_{\perp}-\rangle \langle \nu_{\perp}-\mid G_o\mid \nu_{\perp}-\rangle\nonumber\\
&\times&\langle \nu_{\perp}-\mid H_{-}-H_o\mid \mu_{\perp}-\rangle \langle \mu_{\perp}-\mid G_o\mid \mu_{\perp}-\rangle\nonumber\\
&+&\langle \lambda_{\perp}-\mid G_o\mid \lambda_{\perp}-\rangle\sum_\nu
\langle \lambda_{\perp}-\mid H_{-}-H_o\mid \nu_{\perp}-\rangle \langle \nu_{\perp}-\mid G_o\mid \nu_{\perp}-\rangle\nonumber\\
&\times&\sum_\iota\langle \nu_{\perp}-\mid H_{-}-H_o\mid \iota_{\perp}-\rangle \langle \iota_{\perp}-\mid G_o\mid \iota_{\perp}-\rangle\nonumber\\
&\times&\langle \iota_{\perp}-\mid H_{-}-H_o\mid \mu_{\perp}-\rangle \langle \mu_{\perp}-\mid G_o\mid \mu_{\perp}-\rangle\nonumber\\
&+&\cdots
\end{eqnarray}
}
Assuming short range scattering
\begin{equation}
\langle \iota_{\perp}-\mid H_{-}-H_o\mid \mu_{\perp}-\rangle = U
\end{equation}
the series can be written in the form
{\normalsize
\begin{eqnarray}
\langle \lambda_{\perp}-\mid G\mid \mu_{\perp}-\rangle &=& 
\langle \lambda_{\perp}-\mid G_o\mid \lambda_{\perp}-\rangle 
\langle \mu_{\perp}-\mid G_o\mid \mu_{\perp}-\rangle
\sum_{k=1}^{\infty} U^k \left(\sum_\iota\langle \iota_{\perp}-\mid G_o\mid \iota_{\perp}-\rangle\right)^{k-1}
\end{eqnarray}
}
The infinite series can be formally summed to yield:
\begin{equation}
\langle \lambda_{\perp}-\mid G\mid \mu_{\perp}-\rangle =
\langle \lambda_{\perp}-\mid G_o\mid \lambda_{\perp}-\rangle 
\langle \mu_{\perp}-\mid G_o\mid \mu_{\perp}-\rangle \frac{U}{
1-U\sum_\iota\langle \iota_{\perp}-\mid G_o\mid \iota_{\perp}-\rangle}
\end{equation}
$\sum_\iota\langle \iota_{\perp}-\mid G_o\mid \iota_{\perp}-\rangle$ is directly connected to
the density of on-shell \ess\bs:
\begin{eqnarray}
\sum_\iota\langle \iota_{\perp}-\mid G_o\mid \iota_{\perp}-\rangle
 &=& {\cal P}\sum_{\iota}\frac{1}{E-E_{\iota-}}
-i\pi\sum_{\iota}\delta (E-E_{\iota-})\\
&=&{\cal H} (\Gamma (E)) -i\Gamma (E)
\end{eqnarray}
$\cal P$ and $\cal H$ indicate the principal part and the Hilbert transform, respectively.
$\Gamma (E)$ is defined by:
\begin{eqnarray}
\Gamma (E) &=& \frac{\pi}{\Delta\varepsilon}\;\; {\rm for }\mid E\mid < a\nonumber \\
&=& 0 \;\;\;\;\;\;{\rm  elsewhere}
\end{eqnarray}
$2a$ is the bandwidth of the \ess\bs .

\vspace{.5cm}

\noindent 
The principal value can be estimated by evaluating the integral
\begin{equation}
\int_{-a}^{a}\frac{dE_{\iota-}}{E-E_{\iota -}}
=ln\mid\frac{E+a}{E-a}\mid
\end{equation}
which is zero on-shell and negligible for the \ess near the energy shell.
For the results of the center panel in the second row of fig. \ref{nograv-grav-spec}
the principle part is negligible and $U\sum_\iota\langle \iota_{\perp}-\mid G_o\mid \iota_{\perp}-\rangle$ 
is smaller than unity but of the order of unity:
\begin{equation}
U\sum_\iota\langle \iota_{\perp}-\mid G_o\mid \iota_{\perp}-\rangle = \frac{1}{F}
\end{equation}
where $F$ is a positive number greater unity. Writing
\begin{eqnarray}
\frac{U}{1-\frac{1}{F}}&=&\frac{FU}{F-1}\nonumber\\
&=&\frac{1}{(F-1) \sum_\iota\langle \iota_{\perp}-\mid G_o\mid \iota_{\perp}-\rangle}  \nonumber\\
&\approx &\frac{1}{\sum_\iota\langle \iota_{\perp}-\mid G_o\mid \iota_{\perp}-\rangle}
\end{eqnarray}
the Green function reads:
\begin{equation}
\langle \lambda_{\perp}-\mid G\mid \mu_{\perp}-\rangle = i\frac{\Delta\varepsilon}{\pi}
\langle \lambda_{\perp}-\mid G_o\mid \lambda_{\perp}-\rangle 
\langle \mu_{\perp}-\mid G_o\mid \mu_{\perp}-\rangle 
\end{equation}
Inserting this and eq. (\ref{k-]Hg-]}) in the third term of eq. (\ref{lippmannschwinger2}), yields for this term:
\begin{eqnarray}
\langle g_{\perp}-\mid\tilde{\kappa}_{\perp}-\rangle&\approx&i\frac{\Delta\varepsilon}{\pi}\frac{\Delta V}{NW}\frac{1}{E_{\kappa -}}
\sum_{\lambda ,\mu}E_{\lambda -}
\frac{1}{E_{\lambda -}}\frac{1}{E_{\mu -}}\frac{E_{\mu -}+E_{\kappa -}}{N}\nonumber \\
&=& i\frac{\Delta\varepsilon}{\pi}\frac{\Delta V}{NW}\frac{1}{E_{\kappa -}}\sum_{\lambda ,\mu}\frac{1}{N}\nonumber \\
&=& i\frac{\Delta\varepsilon}{\pi}\frac{\Delta V}{W}\frac{1}{E_{\kappa -}}
\end{eqnarray}
This is of order unity and has the right $E_{\kappa -}$-dependence to yield the telegraph signals.
It is in accordance with the results of the numerical calculations.

\vspace{.5cm}

\noindent 
According to equation (\ref{develop}) 
the time dependent wave packet evolving out of \ga\s is then written as
\begin{eqnarray} \label{wavepacket}
\mid g_\alpha (t)\rangle &=& 
\langle \tilde{\kappa}_{\perp}-\mid g_{\perp}-\rangle\langle g_{\perp}-\mid g-\rangle\langle g-\mid g_\alpha\rangle e^{-iE_{\tilde{\kappa}-}t/\hbar}
\mid \tilde{\kappa}_{\perp}-\rangle\nonumber \\
\end{eqnarray}
Using eqs. (\ref{wkgm}) and (\ref{gmortho}) we obtain
\begin{eqnarray}
\langle g-\mid g_{\perp}-\rangle &=&\sqrt{\frac{NW^2+(\Delta V)^2}{NW^2}}\left( 1-\frac{(\Delta V)^2}{NW^2+(\Delta V)^2}\right)\nonumber \\
&=&\sqrt{\frac{NW^2}{NW^2+(\Delta V)^2}}
\end{eqnarray}
which is of order unity.

\vspace{.5cm}

\noindent 
Defining the on-shell projected state
\begin{equation}
\mid A\kappa -\rangle = \frac{1}{\sqrt{N}}\sum_{\kappa -}\mid\tilde{\kappa}_{\perp}-\rangle
\end{equation}
and projecting eq. (\ref{wavepacket}) on $\mid A\kappa -\rangle$ one obtains:
\begin{eqnarray}
\langle A\kappa -\mid g_\alpha (t)\rangle &=& 
i\frac{\Delta\varepsilon}{\pi}\frac{\Delta V}{W}\sum_{\tilde{\kappa}-}\frac{1}{E_{\kappa -}}
\sqrt{\frac{NW^2}{NW^2+(\Delta V)^2}}
\frac{1}{\sqrt{2}}\frac{1}{\sqrt{N}}e^{-iE_{\tilde{\kappa}-}t/\hbar}   \nonumber \\
&=&
\frac{1}{\pi}\frac{\Delta V}{W}\frac{1}{\sqrt{2N}}
\sqrt{\frac{NW^2}{NW^2+(\Delta V)^2}} \nonumber \\
&\times &\left(\sint +\sinttt +\sinvt +\cdots \right )
\end{eqnarray}
yielding the telegraph signal.
A step-like periodic function $f(x)$, resembling a telegraph signal with 
a constant period, has the following Fourier expansion;
\begin{equation}
f(x)= {\rm sin}x +\frac{{\rm sin}3x}{3}+\frac{{\rm sin}5x}{5}+ \cdots \nonumber
\end{equation}
Hence the superposition of sine-functions with frequencies, which are multiples
of a ground frequency, yields a telegraph-like behaviour.

\vspace{.5cm}

\noindent 
Supporting evidence comes from the spectral distribution of the entanglement of
the initial \p state $\mid g_{\alpha} \rangle$ in the band of \ess\bs
\begin{eqnarray}
\rho_{g\alpha}(E) &=& \sum_{\tilde{\kappa}\pm}\mid\ktpm g_\alpha\rangle \mid^2\delta (E-E_{\kappa\pm}) \\
                     &=& -\frac{1}{\pi}{\rm Im} G_{g\alpha}(E)
\end{eqnarray}
where the Green function $G_{g\alpha}(E)$ is defined by:
\begin{equation}
G_{g\alpha}(E) = \frac{1}{E-E_{g\alpha}-\Sigma_{g\alpha}}
\end{equation}
For energies close to the energy shell the self-energy $\Sigma_{g\alpha}$
is given by:
\begin{eqnarray}
\Sigma_{g\alpha} &=& {\cal P}\sum_{\kappa\pm}\frac{\mid\langle\tilde{\kappa}_{\perp}\pm\mid H\mid g_\alpha\rangle\mid^2}{E-E_{\kappa\pm}}
-i\pi\sum_{\kappa\pm}\mid\langle\tilde{\kappa}_{\perp}\pm \mid H\mid g_\alpha\rangle\mid^2\delta (E-E_{\kappa\pm})
\end{eqnarray}
${\cal P}$ denotes the principal part. The imaginary part of the Green function is:
\begin{equation}
{\rm Im} G_{g\alpha}(E) = \frac{{\rm Im}\Sigma_{g\alpha}}{(E-E_{g\alpha}-{\rm Re}\Sigma_{g\alpha})^2+({\rm Im}\Sigma_{g\alpha})^2}
\end{equation}
The spectral distributions are plotted in fig.  
\ref{nograv-grav-spec} in the panels on the right hand side for the three parameter choices. 
As it was demonstrated in a previous paragraph,
in the telegraph regime (plot in the middle central panel of fig. \ref{nograv-grav-spec}) 
just the anti-symmetric
\p states remain significantly on-shell.
This explains why the spectral weight of 
the initial state on the on-shell eigenstates, derived from the components 
$\{ \mid \kappa- \rangle \}$, is non-zero and 
larger than 
its spectral weight on eigenstates, derived from \ess components $\{ \mid \kappa+ \rangle \}$.
This is obvious from the alternating small and large values
of the spectral weight of $\mid g_{\alpha} \rangle$ on eigenstates.
Therefore, the initial \p state acquires, in its time development, 
\ess components 
with multiple energy differences, hence its time development must resemble
a telegraph signal. 

\vspace{.5cm}              

\noindent 
The resonances in the spectral distributions are of Lorentzian shape.
The time development of the amplitude of the wave packet as given in eq. (\ref{wavepacket})
is directly connected with a sharp resonance of Lorentzian shape.
Such sharp resonances only arise, if the \p state interacts 
very weakly and locally in space with a continuum degenerate in energy.
The locality of the interaction garantees that the coupling does not vary
over the continuum. Only then a Lorentzian shape is obtained.
The weak interaction garantees that the resonance is centered
on the energy shell.

\vspace{.5cm}

\noindent 
\noindent The telegraph signal in our example is a sudden change of state 
of the whole system.
After a time interval suddenly the \p attempts to change side.
Sometimes the attempt is successful, but not every attempt
leads to a side change as it can be seen in the plot in 
fig. \ref{nograv-grav-spec} (middle central panel) at times around 4000 seconds . \\

\vspace{.5cm}

\subsection{\bf Possible application} 
Possible application of our theory is in the field 
of surface phenomena, where telegraph signals have indeed been observed,
associated with state changes in bistable systems.
At low temperature when the diffusion of adsorbates on solid surfaces is
slowed down, the movement of single atoms and molecules can be traced with  
scanning tunnelling microscopy (STM). In these experiments much faster telegraph-signal-like 
jumps of the adsorbate from its adsorption site to a neighbouring one
are observed, compared to the  
time the adsorbate is localized on each adsorption site  (cf. for instance
the hopping time of a hydrogen atom on Cu(100) \cite{lauhon-ho} and of 
one CO molecule in a chevron structure on Cu(111) 
\cite{eiglercascade}). In a current-iduced hydrogen tautomerization reaction 
of naphthalocyanine adsorbed on
a NaCl/Cu(111) surface two hydrogen atoms in the inner cavity change
place reversibly in a telegraph manner at temperature 5K \cite{meyer-telegraph}. 
A telegraph signal is observed by Repp et al. \cite{repp1} as
an adsorbed Au atom on a NaCl/Cu(111) surface switches 
between a higher and a lower oxidation states.  
Current-induced transitions to coverage-dependent excited
states of H$_2$ on Cu(111) at 5K \cite{gupta}, as well as
reversible rotational excitation of O$_2$ on Pt(111) \cite{stipe1}, 
C$_2$HD on Cu(100) \cite{stipe2} and dibutyl sulfide on Cu(111) and Au(111) \cite{sykes,sykes1},  
transition between distinct current states due to conformational
changes \cite{nacci,gaudioso1,gaudioso2,donhauser}, are examples 
of telegraph-signal-like bahaviour of adsorbates in the low temperature STM.
The time oscillations of the Hall photovoltage
accross different pairs of contacts in a Hall bar geometry 
show a statistical telegraph signal \cite{vonklitzing}.

\section{Conclusion}
The quantum theory of the dynamics of a ''particle'' coupling to a continuum
of environmental states,
suggested in the present communication,
yields a telegraph-signal-like time development
of a bistable system as a solution of the
time dependent Schr\"odinder equation.
The coherent time development of the two-state system
entangled to  \ess of different physical
nature reproduces the telegraph behaviour, observed in
experiments of very different kind.
The telegraph signal originates in our theory due to
a weak and local interaction of discrete degrees of freedom with 
on-shell \ess of high density of states. 
The time scale of the telegraph signal is determined by the
energy separation of two adjacent \ess and can vary over a large
range from picoseconds to seconds.

\end{document}